\documentstyle[epsf,epsfig]{elsart}

\newcommand{\nbb}{$0\nu\beta\beta\;$}
\newcommand{\nnbb}{$2\nu\beta\beta\;$}
\newcommand{\x}{$\rm ^{136}Xe\;$}

\begin{document}
\begin{frontmatter}

\title{Detection of very small neutrino masses in double-beta \\
       decay using laser tagging}

\author[ITEP]{M. Danilov,}
\author[IBM]{R. DeVoe,} 
\author[ITEP]{A. Dolgolenko,}
\author[TRIESTE]{G. Giannini,}
\author[STANFORD]{G. Gratta,}
\author[FRASCATI]{P. Picchi,}
\author[UA]{A. Piepke,}
\author[PADOVA]{F. Pietropaolo,}
\author[CALTECH]{P. Vogel,}
\author[NEUCHATEL]{J-L. Vuilleumier,}
\author[STANFORD]{Y-F. Wang,}
\author[ITEP]{O. Zeldovich}

\address[UA]{Physics Department, University of Alabama, Tuscaloosa AL, USA}
\address[CALTECH]{Physics Department, Caltech, Pasadena CA, USA} 
\address[IBM]{Almaden Research Center, IBM, San Jose CA, USA}
\address[ITEP]{ITEP, Moscow, Russia}
\address[FRASCATI]{INFN, Laboratori Nazionali di Frascati, Italy\\
Istituto di Cosmogeofisica del CNR, Torino, Italy\\
Dipartimento di Fisica, Universit\`a di Torino, Italy}
\address[NEUCHATEL]{Institut de Physique, Universite de Neuchatel, Switzerland}
\address[PADOVA]{INFN, Sezione di Padova, via Marzolo 8, Padova, Italy} 
\address[STANFORD]{Physics Department, Stanford University, Stanford CA, USA} 
\address[TRIESTE]{Dipartimento di Fisica, Universit\`a di Trieste, Italy}

\date{\today}
\begin{abstract}
We describe an approach to the study of neutrino masses that
combines quantum optics techniques with radiation detectors 
to obtain unprecedented sensitivity. With it the search for 
Majorana neutrino masses 
down to $\sim$10~meV will become accessible.
The experimental technique uses the  possibility
of individually detecting $\rm Ba^+$-ions in the final state of $\rm ^{136}Xe$
double-beta decay via resonant excitation with a set of lasers aimed at
a specific location in a large Time Projection Chamber.   The specificity
of the atomic levels provides tagging and,
together with more traditional event recognition parameters, greatly 
suppresses radioactive backgrounds.
\end{abstract}


\end{frontmatter}

\section{Introduction}
 
Recent results from a number of independent 
experiments~\cite{RPP} 
can be interpreted as due to finite neutrino masses and, in particular,
high statistics measurements of atmospheric neutrinos by the 
Super-Kamiokande\cite{atm_osc_disc} experiment are regarded by most as 
firm evidence that neutrinos have non-zero masses.\\
While these measurements, based on oscillations, have set the stage 
for a systematic study of the  intrinsic neutrino properties, only upper-limits
exist on the absolute magnitude of neutrino masses.
Indeed, theoretical models span a large range of
scenarios, from the degenerate case where mass differences among flavors
are small with respect to the absolute mass scale~\cite{degenerate}, to the 
hierarchical, where mass differences are of the same order as the mass
themselves.
However, while the neutrino mass scale is unknown, the present data on 
oscillations lead rather naturally to masses in the range 
$0.01 < m_{\nu} < 1$~eV, as shown recently, e.g. in~\cite{bilenki_et_al}.\\
It is unlikely that direct neutrino mass
measurements,  most notably with 
tritium~\cite{tritium},  will be able to
reach sensitivities substantially below 1~eV in the near future.
In contrast, 
we will show that a large double-$\beta$ decay experiment
using isotopically enriched \x can reach a sensitivity corresponding to 
neutrino masses as low as $\sim$0.01~eV.     
A xenon detector offers the unique possibility of identifying the 
final state, thus providing an essentially
background-free measurement of unprecedented sensitivity
for the neutrino mass.

It is well known that neutrinoless double beta decay, \nbb,
can proceed only if neutrinos are massive Majorana particles.  
If the \nbb  occurs,  the 
{\sl effective} Majorana neutrino mass $\langle m_{\nu} \rangle$
is related to the half-life $T_{1/2}^{0\nu\beta\beta}$ as:
\begin{equation}
  \langle m_{\nu} \rangle ^2 = ( T_{1/2}^{0\nu\beta\beta} G^{0\nu\beta\beta}(E_0,Z) 
             |M^{0\nu\beta\beta}_{GT} - {{g_V^2}\over{g_A^2}}M^{0\nu\beta\beta}_F|^2 )^{-1}
  \label{eq:m_t}
\end{equation}
where $G^{0\nu\beta\beta}(E_0,Z)$ is a known phase space factor depending on the 
end-point energy $E_0$ and the nuclear charge $Z$,
and $M^{0\nu\beta\beta}_{GT}$ and $M^{0\nu\beta\beta}_F$ are 
the Gamow-Teller and Fermi nuclear matrix elements for the process.   
Here we have defined
%
$\langle m_{\nu} \rangle = \sum_{i} m_i U^2_{ei}$
%
with $U$ being the mixing matrix in the lepton sector, and $m_i$ the masses of the 
individual Majorana neutrinos.   Hence, although difficulties in the nuclear 
models used to calculate the matrix elements give some uncertainty on the 
value of $\langle m_{\nu} \rangle$ (see, e.g. \cite{petr_new}),
 \nbb decay is sensitive to the masses
of all neutrino flavors, provided that the mixing angles are non-negligible.

\section{Backgrounds and Experimental Limitations}

At present the best sensitivities to \nbb decay have been reached with  
$\rm ^{76}Ge$ diode ionization counters with an exposure of 
24.16~kg~yr~\cite{HeiMos} and with a $\rm ^{136}Xe$ time projection chamber (TPC) 
with an exposure of 4.87~kg~yr~\cite{Xe_gott}. The measured half-life 
limits of $5.7\times 10^{25}$~yrs for germanium and $4.4\times 10^{23}$~yrs 
for xenon can be interpreted as neutrino mass limits of, respectively, 
$\langle m_{\nu} \rangle < 0.2 (0.6)$~eV and 
$\langle m_{\nu} \rangle < 2.2 (5.2)$~eV using the Quasi-Particle Random Phase
Approximation (QRPA)~\cite{QRPA} (Shell Model (NSM)~\cite{SM}) 
for the nuclear matrix element calculation.

The germanium detector rejects background on the basis of its excellent 
energy resolution and of pulse-shape discrimination~\cite{HeiMos} and has
recently reported a specific background as low as 0.3~kg$^{-1}$~yr$^{-1}$~FWHM$^{-1}$.
The relatively inferior energy resolution in TPCs have been complemented 
by superior tracking capabilities that allow the xenon experiment to 
partially reconstruct the two-electron topology of $\beta\beta$-decay,
obtaining a specific background of 2.5~kg$^{-1}$~yr$^{-1}$~FWHM$^{-1}$~\cite{Xe_gott}.
None of these backgrounds is sufficient for a decisive
experiment in the interesting region discussed above. 
For that, it is essential to find a reliable way to drastically 
increase the size of the detectors while, simultaneously, reducing the backgrounds.
This dual requirement stems from the fact that in a background-free experiment 
the neutrino mass sensitivity scales as
%
$\langle m_{\nu} \rangle \propto 1/\sqrt{T^{0\nu\beta\beta}_{1/2}} \propto 1/\sqrt{N t}$,
%
where $t$ is the measurement time and $N$ the number of nuclei under study. 
In the opposite extreme, when the background scales with $Nt$,
the neutrino mass sensitivity would scale only as
%
$\langle m_{\nu} \rangle \propto  1/(N t)^{1/4}$.
%
Obviously, the
backgrounds observed in all current experiments 
will become the true limiting factor of any kind of future large experiment, hampering the full
utilization of very large masses of $\beta\beta$ emitters.
Unlike other isotopes, however, \x allows for direct tagging of the Ba-ion final state
using optical spectroscopy, as pointed-out for the first time in~\cite{moe}.  
While this technique cannot discriminate between \nbb and \nnbb -decays,
this second process is not the dominant background at the sensitivities sought here and
can be eliminated by kinematical reconstruction in a xenon TPC.
Moreover, xenon is one of the easiest isotopes to enrich, and like argon can 
be used as active medium in ionization chambers.    Hence, an experiment with xenon can 
use an entirely new variable in order to drastically reduce the backgrounds and explore the 
range of interesting neutrino masses.

\section{Barium detection in xenon}

The $\beta\beta$-decay of \x produces a Ba$^{++}$ ion in the final state that
can be readily neutralized to Ba$^+$ by an appropriate secondary gas in the TPC. 
The barium ion can then be individually detected through its laser-induced fluorescence.
Single Ba$^+$ ions were first observed in 
1978~\cite{deh2} using a radio frequency quadrupole trap and laser cooling.\\ 
\begin{figure}[htbp]
\begin{center}
\mbox{\epsfig{file=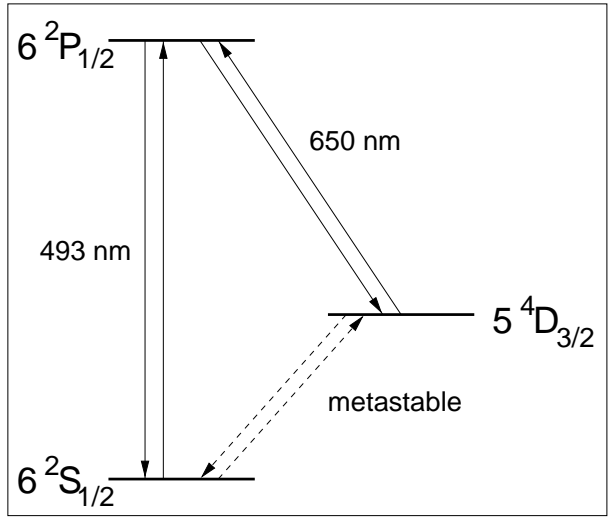,height=8cm,angle=-90,clip=}}
\caption{Atomic level scheme for Ba$^+$ ions.}
\label{fig:ba_levels}
\end{center}
\end{figure}
The level structure of the alkali-like Ba$^+$ ion is shown in 
Figure~\ref{fig:ba_levels}. 
Due to the strong 493~nm allowed transition ground-state ions can be optically 
excited to the $\rm 6^{2}P_{1/2}$ state from where they have substantial branching 
ratio (30\%) to decay into the meta-stable $\rm 5^{4}D_{3/2}$ state.   
Specific Ba$^+$ detection is then achieved by exciting the system back into 
the $\rm 6^{2}P_{1/2}$ state with 650~nm radiation and observing the blue photon 
from the decay to the ground state (70\% branching ratio).    This transition has 
a spontaneous lifetime of 8~ns  and when saturated will radiate 
$6 \times 10^7 $~photons/s.   A pair of lasers tuned onto the appropriate frequencies 
and simultaneously steered to the place where the $\beta\beta$-decay candidate event is found
can provide a very selective validation, effectively providing a new independent
constraint to be used in $\beta\beta$-decay background subtraction.  
The light from the P to S transition 
conveniently lays in the region of maximum quantum efficiency of bialkali photocathodes
so that an array of conventional large-area photomultipliers can be used for the 
detection.   The very large saturation rate makes the experiment possible even with 
modest photocathode angular coverage.   While it is possible in principle to steer the lasers
anywhere inside the TPC, the very large volume and the need for light baffling may favor a
scheme in which the Ba$^+$ drift in the large electric field is used to bring the ion in
specific laser detection regions.

The primary difference between the single atom work done previously and this 
experiment is that here the $\rm Ba^+$ ion is not in vacuum but rather in a buffer 
gas (Xe) at a pressure of several atmospheres.  The high pressure xenon has two effects: 
first, it effectively traps the barium since diffusion in the dense gas is 
sufficient to confine the atoms for long enough time to obtain a signal; second, 
it pressure broadens the optical transitions increasing the laser power required. 
We calculate that at 5~atm the barium ion will diffuse only $\simeq 0.7$~mm in 1~s,
and during this time it can be cycled over $10^7$ times through the three-level scheme
of Figure~\ref{fig:ba_levels}, emitting more than $10^7$ 493~nm photons.
The ion drift in the electric field of the TPC can be accurately measured and corrected
for in the process of steering the lasers.    For the moment we notice that drift 
velocities of Tl$^+$ in Xe have been measured~\cite{Tl-Xe} and are low enough to make the
correction possible. 
We also find from preliminary calculations that at the same pressure the broadened 
line-width is $\sim 20$~GHz, 1000 times greater than the natural line-width. 
This results in a saturation intensity of $\sim 5$~W/cm$^2$.
Ba$^+$ lifetime, together with pressure broadening of the lines involved, and the
lifetime of the $\rm 5^{4}D_{3/2}$ state, will have to be measured in xenon before
the detection system can be fully optimized.
However, measurements on the corresponding neutral Ba states at 0.5~atm in He and 
Ar~\cite{Erlacher} show that operations at 5 to 10~atm should be possible.

\section{A large xenon gas TPC with barium tagging}

Among the different gas ionization detectors known, a TPC is the ideal detector for 
$\beta\beta$-decay as it has no wires or other materials in the detection volume 
hence reducing the background internal to the detector and simplifying the laser scanning.
The requirements of size, efficiency for contained events, energy, and spatial resolution
can be achieved with a large (40~m$^3$) gas-phase TPC with xenon.
TPCs of similar size at atmospheric pressure are successfully in 
operation~\cite{Delphi_TPC,aleph_tpc}, in one case with some 10~kg of \x~\cite{itep}.
A 5 to 10~atm, 40~m$^3$ chamber would contain 
1 to 2~tons of \x and is within the reach of current state-of-the-art technology.

The ultimate sensitivity of the technique is probably limited by the 
practical availability of \x that represents 8.9\% of natural xenon and has to be
extracted by isotopic separation.   While the isotopic separation is indeed one of the 
main challenges of the experiment, in the case of \x this operation is simplified by
the fact that xenon is a gas at standard conditions, and essentially any known
separation technology can be used.     Relatively large amounts of \x have been 
obtained in the past by ultracentrifugation~\cite{roberts} with purities sufficient for
the experiment described~\cite{szady}.    \x handling is simplified by its inert character
that allows for the extraction of cosmogenically produced elements with 
distillation and chemical filtering that can be carried-on in the underground
laboratory prior (and during) the operation of the experiment.
An experiment with up to 10~tons of \x is consistent with the availability of xenon on
the world market~\cite{russian_Xe} and can be setup by assembling a few  TPC
modules inside a single pressure vessel or, possibly, by increasing the pressure 
and/or volume of a single chamber.

\begin{figure}[htbp]
\begin{center}
\mbox{\epsfig{file=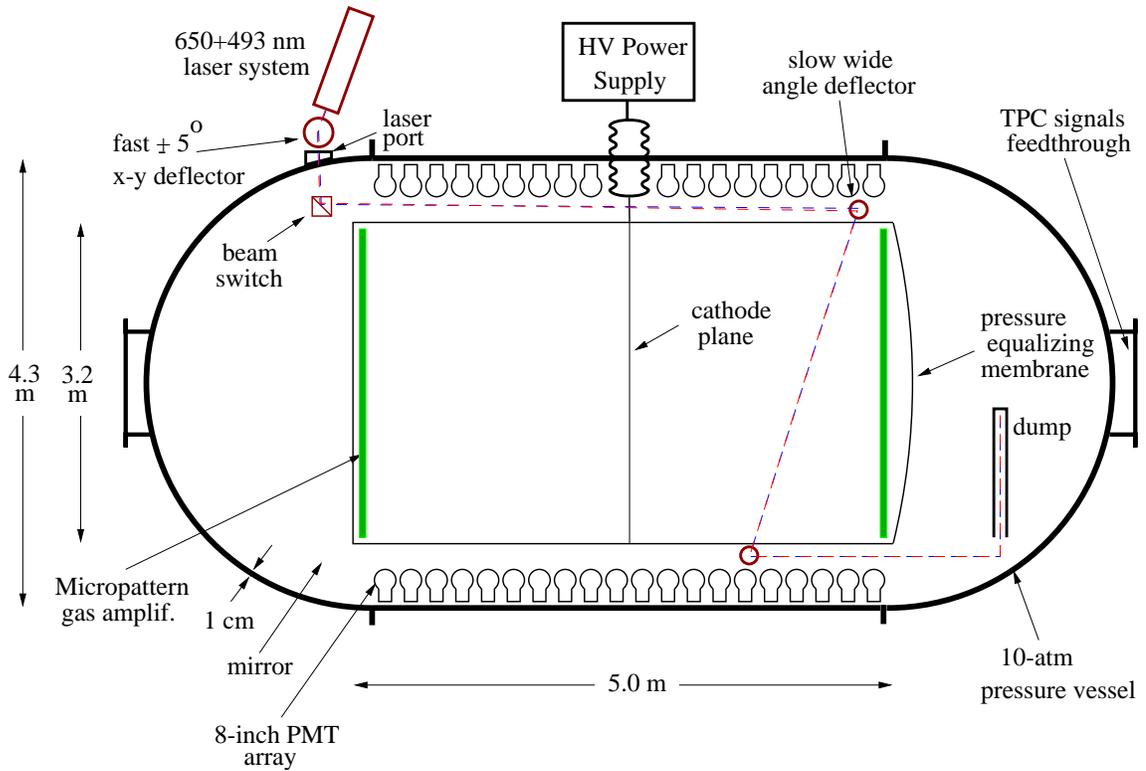,height=15.2cm,angle=-90,clip=}}
\caption{Conceptual layout of the $\beta\beta$ detector.
The active volume of \x is contained by a transparent acrylic cylinder.
An inert buffer gas
transfers the pressure to an outer steel vessel and insulates the
high voltage sections of the TPC. The central
cathode plane is held at a potential of $\approx$250~kV while the
electron multiplication and readout is achieved with micro-pattern
gas amplifier~\cite{GEM} planes. An array of photomultiplier
tubes detects the scintillation light from the xenon and the
fluorescence from the Ba-ions.   All the materials inside the pressure
vessel are selected for low activity.}
\label{fig:TPC_mech}
\end{center}
\end{figure}

The use of liquid-xenon (LXe) in a TPC would result in a very compact detector 
with considerable advantages.  However the range of 1.2~MeV electrons from the 
\nbb -decay in LXe is only of 2.4~mm so that, given any conceivable spatial 
resolution, the topological information, essential for background rejection, would 
be lost.  In contrast, at room temperature 5~atm of xenon gas correspond to a density 
of 30~g/l and an electron range of 21.6~cm at 1.2~MeV.
In Figure~\ref{fig:TPC_mech} we show a possible scheme for the $\beta\beta$-decay
detector.\\
It is known~\cite{wire_chambers} that it is quite difficult to obtain electron
multiplication and stable operation in xenon chambers.  This is due to the far 
UV scintillation light ($\lambda_{\rm scint}\simeq 178$~nm corresponding 
to 6.93~eV), copiously produced with the multiplication process, that
in turn ionizes the gas and extracts electrons from the metallic surfaces.
In our case this particular requirement is greatly relaxed by the use of 
gas micropattern detector~\cite{GEM} planes instead of a more conventional anode wire array for 
the chamber readout, drastically limiting the solid angle available for the UV 
radiation~\cite{pure_argon}.
In addition, the chamber will also use a quencher gas, able to absorb the 
xenon scintillation UV and re-emit light in the blue or green so that timing 
information can be recovered using the photomultipliers and the event can be 
localized also in the third (time) coordinate.
A field of $\sim$1~kV/cm will be needed in order to achieve high drift
velocity for electrons in xenon.   While this field strength is not sufficient 
to neutralize Ba$^{++}$ to Ba$^+$, an appropriate choice of ionization 
potential (IP) for the quencher gas can achieve this purpose.   IPs between 5.2~eV (Ba) and 
10.001~eV (Ba$^+$) are found in several organic molecules that would provide 
a stable environment for Ba$^+$.  Molecules like TMA ($\rm (CH_3)_3N$), 
TEA ($\rm (C_2 H_5)_3 N$), TMG ($\rm (CH_3)_4 Ge$) and TMS ($\rm (CH_3)_4 Si$)
are candidates fulfilling the above requirements.

An underground site with an overburden of $\simeq 2000$~mwe 
(meter water equivalent) will reduce the cosmic ray muon flux to 
$\rm <10^{-2} m^{-2} s^{-1}$, corresponding to less than 0.1~Hz through the detector.  
Muons recorded in the TPC have a distinctive signature and indeed were observed 
and rejected by the Gotthard group~\cite{Xe_gott} based on track length, 
lack of scattering (high energy), and specific ionization. The last two features 
provide good discrimination even for tracks that clip a small corner of the chamber.
The online trigger processor will be able to analyze and reject most muon tracks 
without activating the laser Ba-tagging system. Based on simple geometrical 
considerations we expect the laser system to be activated by muon tracks less than
once per hour.
Some background will be produced by neutrons produced by muon spallation in the
structures (rock and other materials) outside the detector.    These fast neutrons
enter the detector with some efficiency and produce spallation 
reactions on the xenon, carbon and hydrogen nuclei contained in the TPC.
While the original muon and the neutron trail go undetected, the spallation 
processes provide very distinct high ionization, short tracks and can be distinguished
and rejected even at trigger level.
Hence at the depth considered the detector does not need an active veto counter, as
already found by the Gotthard group.\\
More serious are the $\gamma$-ray backgrounds from natural radioactivity, either 
produced outside the TPC (mainly by the rock or concrete), or inside 
(mainly $^{222}$Rn and, possibly, $^{85}$Kr and $^{42}$Ar). 
External $\gamma$-ray backgrounds from the rock and concrete can be attenuated by 
a $\sim$25~cm thick lead or steel enclosure and  by the 1~cm thick pressure vessel 
that will be built out of low-activity steel.
Additionally, cleaner shielding could be provided, if needed. 
In the following we conservatively assume that the TPC itself (without 
Ba-tagging system) will have the same specific rate of mis-identified 
background as the 
Gotthard experiment, and we rely on Ba-tagging for the final step of 
reduction.   This hypothesis is conservative since the larger volume 
of our TPC provides better self shielding for the radiation produced externally.

\section{Discussion}

We estimate the position and energy resolution of our 
chamber from an extrapolation using the Gotthard TPC.
Assuming a drift velocity similar to the one in Gotthard and a drift 
distance of 250~cm (3.6 times larger than the Gotthard case)
we obtain transverse and longitudinal position resolutions ($\sigma$) of 
better than 5~mm.  This figure should give us roughly the same
background suppression factor as obtained at Gotthard, since
the electron range is 21~cm.     Furthermore, unlike in the case of the
Gotthard experiment, the knowledge of the time of occurrence 
will provide  longitudinal localization of each event, improving the 
understanding of external backgrounds and allowing for a longitudinal 
fiducial volume cut.\\
Energy resolution should be no worse than the $\sigma_E/E = 2.5 \%$
(at 1.592~MeV) obtained at Gotthard, since in addition to the total charge,
also the scintillation light will be collected \cite{ion_and_scint}.
The possibility of using the laser tagging system 
with a small spot-size in a ``raster scan'' mode around the event
location, will localize the decay vertex with mm precision and allow
a full kinematical fit including energy, angular
correlation and range for {\it each} of the electrons.
It is expected that substantial improvement in energy resolution
and background rejection will be possible in this way.\\
While the quantitative advantage of each of the techniques discussed 
will be better understood after extensive laboratory tests and a full 
Monte Carlo simulation, here we simply assume that the above methods,
together with the Ba$^+$ identification, will reduce the sum of all
radioactivity backgrounds by
at least three orders of magnitude with respect to the Gotthard case, 
making a 10-ton \x experiment essentially background free.
From the size and geometry of the chamber we obtain an efficiency for fully
contained events of 70\%.
In Table~\ref{tab:performance} we compare the projected sensitivity of 
this experiment with the present limits on \nbb decay.
\footnotesize
\begin{table}[htb!!!!!!!]
\begin{center}
\begin{tabular}{|l|c|c|c|c|c|c|c|c|}
\hline
Isotope and & Total & Enrich. & Det. & Meas. & Bkgd. & $T^{0\nu\beta\beta}_{1/2}$ & \multicolumn{2}{c|}{$\langle m_{\nu}\rangle$} \\
Reference   &  Mass & grade      & eff. &   time      &                             &                            & QRPA                  &             NSM     \\
            &  (kg) & (\%)  & (\%) &  (yr)    &       & (yr)               &    \multicolumn{2}{c|}{(eV)} \\ \hline
$^{76}$Ge~\cite{HeiMos}      & 11   & 86   & 75  & 2.2  & 0.3     & $5.7\times 10^{25}$ & 0.2   & 0.6  \\
\x~\cite{Xe_gott}            & 3.3  & 63   & 22  & 1.47 & 2.5     & $4.4\times 10^{23}$ & 2.2   & 5.2  \\
\x projected                 & 1000 & 65   & 70  & 5    & $0^* +$ & $8.3\times 10^{26}$ & 0.051 & 0.14 \\
(for $\sigma_E / E = 2.8\%$) &      &      &     &      & 1.8 events  &                 &       &      \\    
\x projected                 &10000 & 65   & 70  & 10   & $0^* +$ & $1.3\times 10^{28}$ & 0.013 & 0.037\\
(for $\sigma_E / E = 2.0\%$) &      &      &     &      & 5.5 events  &                 &       &      \\          
\hline
\end{tabular} 
\end{center} 
\caption{\footnotesize Comparison of the best present
double-beta decay experiments and the project described here.  The 
technique of Ba$^+$ tagging will reduce the background, given in units of
kg$^{-1}$~yr$^{-1}$~FWHM$^{-1}$ if not noted otherwise, to the level necessary to fully utilize the
large mass of isotopic species. All other experimental parameters are assumed to be the same 
as in the Gotthard experiment.  We list here, together, the case of an initial detector with
1~ton of \x and the final results possible with 10~tons of \x and a very long (10 years)
data-taking period. The quantities marked with $^*$ are radioactivity
backgrounds that are assumed to be negligible as discussed in the text. In addition the 
background from mis-identified \nnbb decays is also shown in total events in each exposure
using the asymmetric analysis interval described in the text.}
\label{tab:performance}
\end{table}
\normalsize

It is interesting to note that for the very large isotopic masses contemplated
here, the background on \nbb decay from the well known \nnbb mode has to be carefully
estimated.    As already remarked this ``background'' is not directly suppressed 
by the laser tagging methods.   However, the energy spectra of the electrons that
represent the only distinctive feature of this background 
will be better measured
thanks to the better knowledge of the event kinematics.   In the following we will
conservatively disregard this additional information and suppress the \nnbb mode 
using total energy information, with the resolution mentioned above.   
In order to understand the role of the resolution we select events in the two 
intervals $I_{sym} = [Q_{\beta\beta}-2\sigma_E,Q_{\beta\beta}+2\sigma_E]$ and
$I_{+} = [Q_{\beta\beta},Q_{\beta\beta}+2\sigma_E]$.
We then compute the number of \nnbb decay events~\cite{petr_new} left in each case.
For $\sigma_E / E = 2.8\%$ we have 23~$\rm events~yr^{-1}ton^{-1}$ left in $I_{sym}$
and 0.36~$\rm events~yr^{-1}ton^{-1}$ left in $I_{+}$. 
A better resolution of $\sigma_E / E = 2\%$ is used in the 10~ton~-~10~yr exposure in
Table~\ref{tab:performance}.
The loss in efficiency due to the asymmetric cut (and to the tails beyond $2\sigma$)
are taken into account in the table as appropriate.    It is clear, however, that our
estimate is quite conservative as the \nnbb decay background will be further suppressed
by the kinematic fitting.

The projected neutrino mass sensitivity of 10 - 50~meV would make the discussed 
experiment competitive with other large scale double beta searches which 
have been proposed. 
NEMO3 is scheduled to start data taking in 2001 with initially 7~kg of $^{100}$Mo 
and 1~kg of $^{82}$Se in form of passive source foils in a gas tracking detector 
utilizing a magnetic field and a scintillator calorimeter~\cite{nemo3}.
A neutrino mass sensitivity of 0.1~eV is expected after 5 years. The ultimate goal
is to run with 20 kg of $^{82}$Se, $^{100}$Mo or $^{150}$Nd in order to measure an
effective neutrino mass below 0.1~eV. 
A large cryogenic detector (CUORE), able to operate 600 kg of
TeO$_2$ crystals, could be used to search for the \nbb decay of $\rm ^{130}Te$. 
A neutrino mass sensitivity of around 0.1 eV has been quoted for this device
\cite{fiorini}. 
The proponents of the GENIUS project propose to use up to one ton of isotopically
enriched $^{76}$Ge suspended in a large tank of liquid nitrogen \cite{genius}.
The goal is to reach a mass sensitivity of around 0.01 eV.
The latter two are calorimetric approaches.
However, it has to be pointed out that the experiment discussed in this paper is the
only one among those third generation projects which plans to utilize a novel
technique for background suppression in the form of Ba tagging.

In summary, we have described an advanced $\beta\beta$ detector system that uses
a hybrid of atomic and particle physics techniques to qualitatively improve
background suppression.      Such detector opens new possibilities in using massive
amounts of \x for an advanced $\beta\beta$-decay experiment that would explore neutrino 
masses in the range 10 - 50~meV, providing a unique opportunity for discoveries in particle
physics and cosmology.

\section{Acknowledgments}

We would like to specially thank F. Boehm for the help and guidance
received in formulating much of this paper.
We also owe gratitude to U.~Becker, S.~Chu, H.~Henrikson, 
L.~Ropelewski, T.~Thurston and R. Zare for many useful discussions.
Finally we would like to thank R.G.H.~Robertson for pointing out an inconsistency in the 
original version of Table~\ref{tab:performance}.

\newpage
%
%
\newpage

%
%


\begin{thebibliography}{99}


\bibitem{RPP} See e.g. Neutrino section in C.~Caso et al., Eur. Phys. J. C 3 (1998) 1.

\bibitem{atm_osc_disc} Y. Fukuda et al., 
                       Phys. Rev. Lett. 81 (1998) 1562.

\bibitem{degenerate} D.O.~Caldwell and R.N.~Mohapatra, Phys. Rev. D 48 (1993) 3259;\\
                  J.~Ellis and S.~Lola, Phys. Lett. B 458 (1999) 310;\\
                  V.~Barger and K.~Whisnat, Phys. Lett. B 456 (1999) 194.



\bibitem{bilenki_et_al} S.M. Bilenkii et al., Phys. Lett. B 465 (1999) 193.

\bibitem{tritium} P.~Fisher, B.~Kayser and K.~McFarland, 
                  Ann. Rev. Nucl. Part. Sci. 49 (1999) 481.

\bibitem{petr_new} P.~Vogel to appear in ``Current aspects of Neutrino Physics'' 
                  D.O.~Caldwell ed., Springer, to appear in 2000.

\bibitem{HeiMos}  L. Baudis et al., Phys. Rev. Lett. 83 (1999) 41.

\bibitem{Xe_gott} R. Luescher et al., Phys. Lett. B 434 (1998) 407.

\bibitem{QRPA} A. Staudt et al., Europhys. Lett. 13 (1990) 31.
 
\bibitem{SM} E. Caurier et al., Phys. Rev. Lett. 77 (1996) 1954.


\bibitem{moe} M.K. Moe, Phys. Rev. C 44 (1991) R931.

\bibitem{deh2} W. Neuhauser, M. Hohenstatt, P. Toschek, and H. Dehmelt, 
               Phys. Rev. Lett. 41 (1978) 233.

\bibitem{Tl-Xe} T.M. Maddern and L.W. Mitchell, Nucl. Inst. and Meth. A 359 (1995) 506.

\bibitem{Erlacher} E. Erlacher and J. Hunnekens, Phys. Rev. A 46 (1992) 2642 (1992).

\bibitem{Delphi_TPC} C. Brand et al., Nucl. Inst. and Meth. A 283 (1989) 567.

\bibitem{aleph_tpc} W.B. Atwood et al., Nucl. Inst. Meth. A 306 (1991) 446.

\bibitem{itep} V.A.~Artemiev et al.,
                Nucl. Inst. and Meth. A303 (1991) 309;\\
		V.~Artemiev et al., Phys. Lett. B 345 (1995) 564.

\bibitem{roberts} W.L. Roberts Nucl. Inst. and Meth. A 282 (1989) 271.

\bibitem{szady} E. Bellotti et al. Nucl. Inst. and Meth. B 62 (1992) 529.

\bibitem{russian_Xe} P.K. Lebedev and V.I. Pryanichnikov, Nucl. Inst. and Meth. A 327 (1993) 222.

\bibitem{wire_chambers} F. Sauli, CERN Yellow Report 77-09, Geneva 3 May 1977.

\bibitem{GEM} For a general review on gas micro-pattern detectors see:
                F. Sauli and A. Sharma, Ann. Rev. Nucl. Part. Sci. 49 (1999) 341.

\bibitem{pure_argon} A. Bressani et al., Budker Institute Preprint INP-98-59,
                     Novosibirsk 1998.

\bibitem{ion_and_scint} J. S\'eguinot et al., Nucl. Inst. and Meth. A 354 (1995) 280.

\bibitem{nemo3} NEMO3 proposal, LAL preprint 94-29 (1994);\\
      C. Marquet for the NEMO coll., to appear in Nucl. Phys. B (Proc.Suppl) 87 (2000).

\bibitem{fiorini} E. Fiorini, Phys. Rep. 307 (1998) 309.

\bibitem{genius} L. Baudis et al., Phys. Rep. 307 (1998) 301.






\end{thebibliography}
\end{document}